\documentclass{elsart}

\usepackage{natbib}

\newcommand{\nsteps}{\mbox{$n_{\rm steps}$}}
\newcommand {\apgt} {\ {\raise-.5ex\hbox{$\buildrel>\over\sim$}}\ }
\newcommand {\aplt} {\ {\raise-.5ex\hbox{$\buildrel<\over\sim$}}\ }

\usepackage{graphicx}

\usepackage{amssymb}

\begin{document}

\begin{frontmatter}



\title{Distributed N-body Simulation on the Grid Using Dedicated Hardware}


\author[aff:scs,aff:api]{Derek Groen}, 
\author[aff:scs,aff:api]{Simon Portegies Zwart}, 
\author[aff:dru]        {Steve McMillan},
\author[aff:utk]        {Jun Makino}

\address[aff:scs]{
  Section Computational Science, University of Amsterdam, Amsterdam, the Netherlands
} 
\address[aff:api]{
  Astronomical Institute "Anton Pannekoek", University of Amsterdam, Amsterdam, the Netherlands
}
\address[aff:dru]{
  Drexel University, Philadelphia, United States
}
\address[aff:utk]{
  University of Tokyo, Tokyo, Japan  
}

\begin{abstract}
We present performance measurements of direct gravitational $N$-body
simulation on the grid, with and without specialized (GRAPE-6)
hardware.  Our inter-continental virtual organization consists of
three sites, one in Tokyo, one in Philadelphia and one in
Amsterdam. We run simulations with up to 196608 particles for a
variety of topologies. In many cases, high performance simulations
over the entire planet are dominated by network bandwidth rather than
latency.
With this global grid of GRAPEs our calculation time remains dominated
by communication over the entire range of $N$, which was limited due
to the use of three sites. Increasing the number of particles will
result in a more efficient execution.
Based on these timings we construct and calibrate a model to predict
the performance of our simulation on any grid infrastructure with or
without GRAPE.  We apply this model to predict the simulation
performance on the Netherlands DAS-3 wide area computer.  Equipping
the DAS-3 with GRAPE-6Af hardware would achieve break-even between
calculation and communication at a few million particles, resulting in
a compute time of just over ten hours for 1~$N$-body time unit.

\end{abstract}


\end{frontmatter}

\section{Introduction}
Star clusters are often simulated by means of direct-method $N$-body
simulations \citep{1985IAUS..113..251A}.  The Newtonian gravitational force on
individual stars in such simulations is calculated by aggregating the
force contributions from all other particles in the system.

To enable faster execution of these simulations, specialized solutions
such as GRAvity PipEs (GRAPEs) \citep{2005PASJ...57.1009F}, Graphics
Processing Units (GPUs)
\citep{2007astro.ph..2058P,2007astro.ph..3100H,2007arXiv0707.0438B}
and Field-Programmable Gate Arrays (FPGAs)
\citep{10.1109/FPGA.2002.1106673} have been successfully developed and
applied. These solutions are designed or tuned specifically for
optimizing force calculations, and provide dramatic speedup. For
example, the GRAPE-6Af features a dedicated hardware implementation
that can calculate 42 force interactions simultaneously and with
increased efficiency. As a result, the GRAPE is able to perform force
calculations $\sim 130$ times faster than a single PC
\citep{2003PASJ...55.1163M}. Recently, GPUs have shown gains in speed and
flexibility, and they are now used for simulating self gravitating systems
at speeds comparable to GRAPE
\citep{2007astro.ph..2058P,2007astro.ph..3100H,2007arXiv0707.0438B}.

Parallelization of GRAPEs appears to be an efficient way to reduce the
wall-clock time for individual simulations
\citep{10.1109/SC.2003.10021,DBLP:journals/pc/GualandrisZT07,2007NewA...12..357H}. The
gravitational $N$-body problem has calculation time complexity
$\mathcal{O}(N^2)$, whereas the communication scales only with
$\mathcal{O}(N)$. For sufficiently large $N$, the force calculation
time will therefore overtake the communication time.  For a local cluster of
GRAPEs with low-latency and high bandwidth network, break-even between
calculation and communication is reached at $N \sim 10^4$
\citep{2007NewA...12..357H}.

Generally, GRAPE clusters are not cheap and few institutions can
afford such dedicated hardware solutions. Still, more than 500 GRAPE
modules, where one module is equivalent to one GRAPE-6Af, or 4 GRAPE-6
chips, are currently in use across 37 institutions in 12 countries
world-wide. An alternative to purchasing a large GRAPE-6 or GPU
cluster is provided by a computational grid. In a grid, several
institutions assemble in a virtual organization, within which they
share resources, and the costs for purchasing and maintaining these
resources \citep{IanFoster08012001}. Grid middleware provides a
secure wide area computing environment without requiring users to
register for individual clusters. In addition, grid-enabled MPI
implementations, such as MPICH-G2 \citep{2002cs........6040K} or
OpenMPI \citep{richard06:_open_mpi}, provide the ability to run MPI
jobs across sites in the grid, using the existing MPI
standards. Applying such grid technology to clusters of GPUs is an 
attractive option, because there
are a large number of (frequently idle) GPUs in consumer machines. By
connecting these consumer machines to the grid (as was done in a
similar fashion with regular CPUs for the SETI@home project
\citep{581573}) and using them for parallel N-body simulations, we can
increase the computational power of the grid in a cheap and convenient
manner.

Although there is a clear benefit of using grid technology in sharing
financial burden, the real challenge is to develop new applications
for astronomical problems that have yet to be solved. For example, the
simulation of an entire galaxy, requires at least a few PFLOP/s of
computational power and the development of a hybrid simulation
environment \citep{2007astro.ph..3485H}. Such an environment performs
several astrophysical simulations on vastly different temporal and
spatial scales. For example, a hybrid simulation environment could
consist of a stellar evolution simulation to track how individual 
stars evolve over time, a smoothed particle hydrodynamics simulation
\citep{1992ARA&A..30..543M} to simulate stellar collisions or close
encounters, and a direct-method N-body calculation to 
simulate the remaining dynamics between stars.

To facilitate these tightly-coupled multi-physics simulations on the
PFLOP/s scale, it will no longer be sufficient to do high-performance
computing (HPC) on a local cluster, as we require an extensive grid
infrastructure consisting of several of such clusters.  Although grid
technology has been largely applied to facilitate high-throughput
computing \citep{10.1109/IPDPS.2000.846030}, little research has been
done on investigating how the grid can be efficiently applied to solve
tightly-coupled HPC problems. By using grid technology for this
specific set of problems, we can potentially fullfill the
computational requirements for performing petascale multi-physics
simulations.

Using a grid infrastructure for HPC has drawback, as the
communication between grid sites dramatically increases network
overhead compared to a local cluster.  For intercontinental
communication, the network latency can become as large as 0.3s, which
is especially impractical for applications, such as direct-method
N-body codes, that require communication over all processes during
every iteration.  Still, even for such long communication paths there
will be a problem size ($N$) for which wall-clock time is dominated by
the force calculation rather than by communication.  Earlier
experiments indicate that a grid of regular PCs across Europe improves
overall performance for relatively small $N$
\citep{DBLP:journals/pc/GualandrisZT07}. We address the question for
which problem size a world-wide grid has practical usage, in
particular if such a cluster is equipped with GPUs or GRAPEs.

\section{Experimental setup}\label{Sect:setup}

We have constructed a heterogeneous grid of GRAPEs, which we call the
Global GRAPE Grid (or G3). The G3 consists of five nodes across three
sites.  Two nodes are located at Tokyo University (Tokyo, Japan), two
are located at the University of Amsterdam (Amsterdam, the
Netherlands) and one is at Drexel University (Philadelphia, United
States).  Each of the nodes is equipped with a GRAPE-6Af special
purpose computer, which allows us to test several different resource
topologies. Local nodes are connected by Gigabit Ethernet, whereas
the different sites are connected with regular internet.  In
Table~\ref{Tab:G3} we present the specifications of the G3. Each of
the computers in the G3 is set up with Globus Toolkit
middleware\footnote{http://www.globus.org} and
MPICH-G2\footnote{http://www3.niu.edu/mpi/, in the future:
http://dev.globus.org/wiki/MPICH-G2}.

\begin{table}
\caption{Specifications for the nodes in G3.
         The first column gives the name of the computer followed by
         its country of residence (NL for the Netherlands, JP for
         Japan and US for the United States). The subsequent columns
         give the type of processor in the node, the amount of RAM,
         followed by the operating system, the kernel version and the
         version of Globus installed on the PC. Each of the nodes is
         equipped with a 1\,Gbit/s Ethernet card and GRAPE-6Af hardware. 
         Local nodes are interconnected with Gigabit Ethernet.}
\begin{tabular}{lclrrrrr}
\hline
name     &location&CPU type    &RAM   &  OS      &kernel      & Globus\\
         &        &            &[MB]  &          &\multicolumn{2}{c}{version}\\
\hline
vader    & NL &Intel P4 2.4GHz & 1280 &Ubuntu 5.10   & 2.6.5   & 4.0.3 \\
palpatine& NL &Intel P4 2.67GHz&  256 &RHEL 3       & 2.4.21  & 4.0.3 \\
yoda     & JP &Athlon 64 3500+ & 1024 &FC 2     & 2.6.10  & 3.2.1 \\
skywalker& JP &Athlon 64 3500+ & 1024 &FC 2     & 2.6.10  & 3.2.1 \\
obi-wan  & US &2x Xeon 3.6GHz  & 2048 &Gentoo 06.1  & 2.6.13  & 4.0.4 \\
\hline
\label{Tab:G3}
\end{tabular}
\end{table}

In Table~\ref{Tab:Network} we present the network characteristics,
latency and bandwidth, of the connections within G3. We tested local
area network (LAN) and wide area network (WAN) connections using the
UNIX {\tt ping} command to measure latency. We use {\tt scp} for
measuring the network bandwidth, transferring a 75 MB file, rather
than referring to theoretical limits because the majority of bandwidth
on non-dedicated WANs is used by external users. For our performance
measurements, we used a standard implementation of MPICH-G2 without
specific optimizations for long-distance networking. As a result, the
MPI communication makes use of only 40\%-50\% of the available bandwidth
\footnote{for more information we refer to a research report from
INRIA: http://hal.inria.fr/inria-00149411/en/}. If we were to enhance
MPICH-G2 with additional optimizations, or add support for grid 
security to already optimized MPI libraries, such as Makino's 
tcplib\footnote{see:http://grape.mtk.nao.ac.jp/$\sim$makino/softwares}
or OpenMPI, our bandwidth use would be close to the bandwidth use of a 
regular file transfer.

\begin{table}
\caption{Characteristics of local and wide network
         connections. Latency indicates the required time for sending
         1 byte through the network connection. The bandwidth
         indicates the transfer capacity of the network
         connection. The bandwidth was measured with a 75MB {\tt scp} file
         transfer.}
\begin{tabular}{lr@{.}lr@{.}lr@{.}l}
\hline
connection &\multicolumn{2}{c}{latency}&\multicolumn{2}{c}{bandwidth (theory)}&\multicolumn{2}{c}{bandwidth (real)}\\
&\multicolumn{2}{c}{[ms]}&\multicolumn{2}{l}{[MB/s]}&\multicolumn{2}{l}{[MB/s]}\\
\hline
Amsterdam LAN               &  0&17  &125&0 & 11&0  \\
Tokyo LAN                   &  0&04  &125&0 & 33&0  \\
Amsterdam - Tokyo WAN       &266&0   &57&0  &  0&22 \\
Amsterdam - Phil. WAN&104&0   &312&5 &  0&56 \\
Philadelphia - Tokyo WAN    &188&0   &57&0  &  0&32 \\
\hline
\label{Tab:Network}
\end{tabular}
\end{table}

The N-body integrator we have chosen for our experiments uses block
time-steps \citep{1986LNP...267.....H} with a 4th order Hermite
integration scheme \citep{1992PASJ...44..141M}.  The time steps with
which the particles are integrated are blocked in powers of two
between a minimum of $2^{-22}$ and a maximum of $2^{-3}$. During each
time step, the codes perform particle predictions, calculate forces
between particles and correct particles on a block of active
particles. Particle corrections include updates of positions and
velocities, and computation of new block time steps of particles. For
our experiments we use three implementations of a parallel $N$-body
integrator. One of these codes runs on a single PC with and without
GRAPE. The two others are parallelized using MPI: one of these uses
the copy algorithm \citep{2002NewA....7..373M,775815} and the other
uses the ring algorithm
\citep{43389,83774,DBLP:journals/pc/GualandrisZT07}. The copy
algorithm has smaller number of communication steps whereas the ring
algorithm has lower memory usage on the nodes.

We initialize the simulations using Plummer
\citep{1911MNRAS..71..460P} spheres that were in virial
equilibrium and performed our simulations using a softening parameter 
of $2^{-8}$. Since our simulations are performed over one dynamical
($N$-body) time unit \citep{1986LNP...267..233H}, the realization of
the N-body system is not critical to the timing results.

\section{Results of grid calculations}\label{Sect:Results}

We have performed a number of simulations on local machines and on the
G3, which consists of simulations lasting one N-body time unit and shorter 
simulations lasting one integration time step. We measured the full 
wall-clock execution time for the longer simulations and we profiled 
the shorter simulations.

\subsection{Timing results of N-body calculations}

\begin{figure}
  \centering \includegraphics[angle=270,scale=0.4]{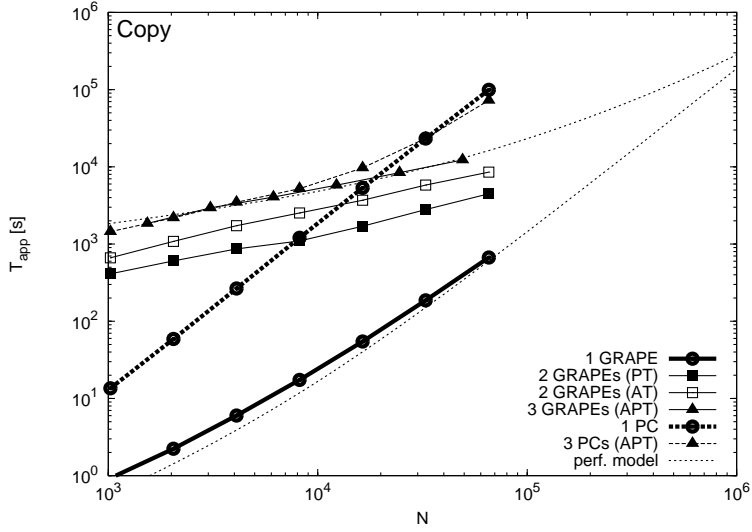}
  \caption{The time for running the application for 1 $N$-body time
           unit ($T_{\rm app}$) as a function of the number of stars
           ($N$) using the copy algorithm. The two thick lines give
           the results for a single CPU with GRAPE (lower solid curve)
           and without (top dashed curve). We make the general
           distinction between solid curves to present the results for
           simulations run with GRAPE, and dashed curves to give the
           results without GRAPE. The results on the grid are
           presented with four different lines, based on the three
           included locations. Each of these runs is performed with
           one node per site. The results for the WAN connection
           Philadelphia--Tokyo, Amsterdam--Tokyo and
           Amsterdam--Philadelphia--Tokyo are indicated with the solid
           curves with filled squares, open squares and filled
           triangles, respectively. The dashed curve with filled
           triangles give the results for the
           Amsterdam--Philadelphia--Tokyo connection but without using
           GRAPE. Dotted lines indicate the performance of runs with
           GRAPE according to the performance model.}
\label{Fig:Copy}
\end{figure}

\begin{figure}
  \centering
  \includegraphics[angle=270,scale=0.4]{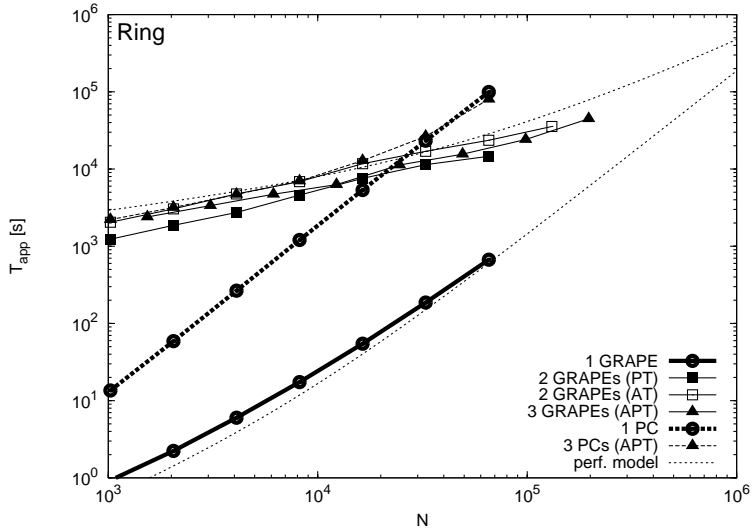}    
  \caption{The time for running the application for 1 $N$-body time unit
           ($T_{\rm app}$) as a function of the number of stars ($N$)
           for runs using the ring algorithm. See Fig.1 for an
           explanation of the lines and symbols.}
\label{Fig:Ring}
\end{figure}

We run the $N$-body codes, discussed in \S\,\ref{Sect:setup}, on a
single PC and across the network in parallel using $N=1024$ to
$N=65536$ (a few additional calculations were performed with
$N>65536$). The runs were performed with and without GRAPE. In
Figs.\,\ref{Fig:Copy} and \ref{Fig:Ring} we present the results of the
copy and ring algorithms. If a simulation is run multiple times with
the same problem set, the execution time may be slightly different per
run. This variation is relatively small, as the slowest of 4 repeated
runs (using 32768 particles over two sites) was a factor 1.07 slower
than the fastest run. The variation can be primarily
attributed to fluctuations in the network bandwidth.

\textbf{Single PC}\\ The performance on a single PC (represented by
the thick solid line with bullets in Fig.\ref{Fig:Copy}) is entirely
dominated by force calculations, which scales as
$\mathcal{O}(N^2)$. As the number of steps per N-body time unit
increases with $N$, the execution time scales slightly worse than
$N^2$.

\textbf{Grid of PCs}\\ The performance on the G3, without using GRAPE,
is given by the thin dashed line with triangles. For $N<24576$, the
performance is dominated by network communication. Given that $p$
indicates the number of processes, the network communication scales as
$\mathcal{O}\left(N \log{p}\right)$(see
\citep{2007NewA...12..357H}). For our grid-based simulation
experiments without GRAPE, break-even between communications and force
calculations is achieved around $N\sim 3 \cdot 10^4$ for the copy
algorithm (Fig.\ref{Fig:Copy}), and at a somewhat higher value for the
ring algorithm (Fig.\ref{Fig:Ring}). For larger $N$, the execution
time is dominated by force calculations, rather than network
communication. For these high $N$, the grid speedup \emph{$\Gamma$}
\citep{DBLP:conf/egc/HoekstraS05}, which is the single-site execution
time divided by the execution time over three grid sites, increases to
$1.37$ for the copy and $1.24$ for the ring algorithm. As can be seen
by comparing Figs.\,\ref{Fig:Copy} and \ref{Fig:Ring}, the copy
algorithm gives overall better performance than the ring
algorithm. This can be explained by the smaller number of
communication steps in the copy algorithm.

\textbf{Single PC with GRAPE}\\ The performance on a single PC with
GRAPE is dominated by force calculations, although communication
between host and GRAPE, and operations on the host machine have an
impact on performance for $N<16384$. In addition, the GRAPE performs
less efficiently for low $N$, because many blocks are too small to
fill the GRAPE pipelines. For larger $N$, force calculations become
the performance bottleneck, and the scaling of the execution time
becomes that of a single PC without GRAPE.

\textbf{Grid of PCs with GRAPE}\\ The performance on the G3 (with
GRAPEs) using all three sites is given by the thin solid line with
triangles. For all problem sizes $N$ we have measured, the grid
speedup \emph{$\Gamma$} is less than $0.15$, indicating that the
performance is dominated by network communication. The network
communication time scales better than the force calculation time,
therefore, force calculation time will overtake the network
communication time if $N$ is sufficiently large. However, this
break-even point lies at much higher $N$ than for a Grid of PCs,
because the use of GRAPE greatly decreases the time spent on force
calculations.

For the copy algorithm (see Fig. \ref{Fig:Copy}), calculations between
Tokyo and Philadelphia take less time than calculations between
Amsterdam and Tokyo, due to a lower network latency (see Table
\,\ref{Tab:Network}). The calculations across three sites take more
time than calculations across two sites. This is caused by the latency
of all-to-all MPI communications in the copy algorithm, which scales
with the number of processors.

According to our profiling measurements in Fig. \ref{Fig:analysis},
for $N<12288$, a simulation on the G3 with GRAPEs using ring algorithm
spends most of its time in network latency. For larger $N$
more time is spent on using the network bandwidth. These results
indicate that network bandwidth is the primary bottleneck for our
simulations on the G3 using ring algorithm. When we compare the
results of the runs on the grid with GRAPEs with each other, we do not
notice any systematic trend. The results confirm that the wall-clock
time is dominated by using the network bandwidth, which is
bottlenecked by the transpacific network line for all grid setups.

\subsection{Profiling of the N-body simulations}

\begin{figure}
  \centering
  \includegraphics[angle=270,scale=0.4]{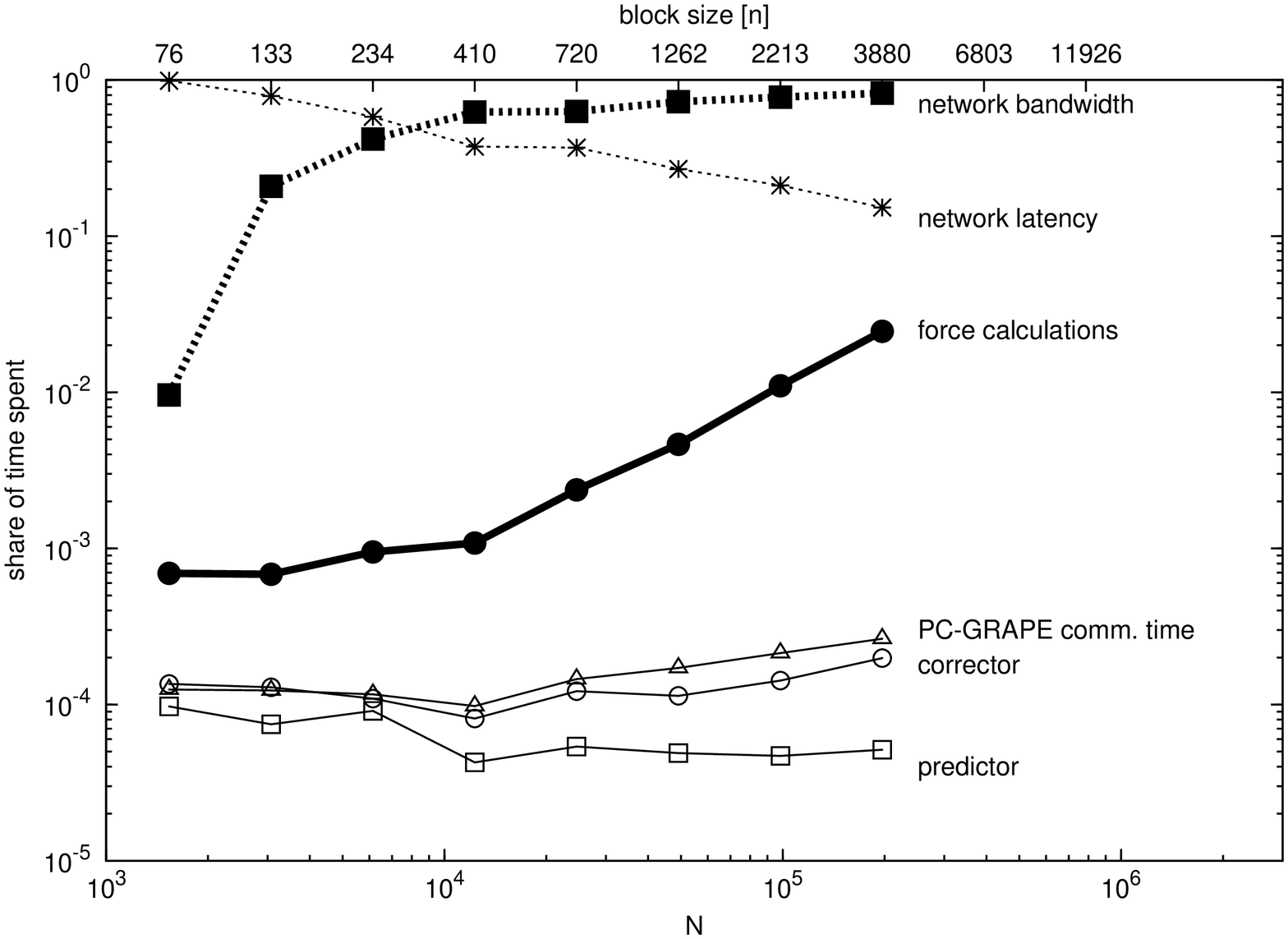}    
  \caption{ Share of wall-clock time spent on individual tasks during a single
time-step. Solid lines indicate tasks performed on the local machine.
The thick solid line with filled circles represents time spent on
force calculations, and the thin solid lines give the result for time
spent on communication between PC and GRAPE (open triangles), particle
corrections (open circles) and particle predictions (open squares)
respectively. Dotted lines indicate time spent on communication
between nodes. The thin dotted line with asterisks indicates time
spent on communication latency between nodes and the thick
dotted line with solid squares indicates time spent on using the
network bandwidth.
  \label{Fig:analysis}
  }
\end{figure}

We have chosen one parallel algorithm (ring) and one resource topology
(3 nodes on 3 sites) to profile the simulation during one integration
time step. The block size $n$ for every measurement was fixed using a
formula for calculating average block size $\left(n = 0.20
N^{0.81}\right)$, which has been used for the same initial conditions
in \citet{2007astro.ph..2058P}. During execution, we measured the time
spent on individual tasks, such as force calculations or
communication latency between processes. We have profiled our
simulations for $N=1024$ up to $N=196608$, using the timings measured
on the process running in Tokyo. The results of these measurements are given in Fig.
\ref{Fig:analysis}. 

We find that for larger $N$, low bandwidth of our
wide area network affects the outcome of the performance
measurements, and that MPI calls are only able to use about a quarter
of the available bandwidth for passing message content. For $N \apgt 5
\cdot 10^{5}$ we expect the force calculation to take more time than
network latency. If we were to use the 
network bandwidth more efficiently for such a large number of particles, 
the execution time would be dominated by force calculations. The network bandwidth 
can be used much more efficiently, either by using a more efficient MPI 
implementation (e.g. one that supports communication over multiple 
tcp connections) or by using a dedicated network.
Using our current networking and MPI implementation, we expect that for $N \apgt 2 \cdot 10^{6}$ particles the force 
calculation time overtakes the bandwidth time.



\section{Modelling the performance of the grid}\label{Sect:model}

In order to further understand the results and to enable
performance predictions for larger network setups, we decided to model 
the performance of the grid calculations.
We model the performance of the simulation by adopting the parallel
performance models described by \citet{2002NewA....7..373M} and
\citet{2007NewA...12..357H} and combining it with the
grid performance model described in
\citet{DBLP:journals/pc/GualandrisZT07}. Further extension and calibration
of the model allows us to simulate the performance of our N-body simulations
on a G3 or any other topology.

\subsection{Single PC}\label{Sect:Sequential}


An $N$-body simulation over one $N$-body time
unit \citep{1986LNP...267..233H} consists of the following steps:
\begin{enumerate}
\item Read the input snapshot and initialize the N-body system.
\item Compute the next system time $t$ and select the block of $n$ active particles in the system.
\item Predict the positions and velocities of all $N$ particles to time $t$.
\item Calculate the forces and their time derivatives between the $n$ active particles and all $N$ particles in the system. 
\item Correct the positions, velocities and velocity derivatives of the $n$ active particles, and update their time steps.
\item Repeat from step 2 until $t$ has exceeded one N-body time unit.
\item Write the output of the simulation and terminate it.
\end{enumerate}

As relatively little time is spent on program initialization and finalization, we focus on the time to
integrate the system ($T_{\rm integrate}$), which consists of the tasks performed in steps 2 to 5. Throughout this paper we use uppercase $T$ to refer to the time spent in $\nsteps$ integration steps, and the lowercase ($t$) for the time spent in a single step. Within a single step, the total execution time $T_{\rm integrate}$ is
\begin{eqnarray}
  T_{\rm integrate} &=& \sum^{\nsteps}_{i=1} 
                      \left( t_{\rm pred} + t_{\rm force} + t_{\rm corr}
		      \right),
\label{Eq:Tintegrate}
\end{eqnarray}
with the time spent on predicting particles
\begin{eqnarray}
  t_{\rm pred} &=& \tau_{\rm pred} N,
\end{eqnarray}
the time spent on calculating forces
\begin{eqnarray}
  t_{\rm force} &=& \tau_{\rm force} n N,
\end{eqnarray}
and the time spent on correcting the active particles
\begin{eqnarray}
  t_{\rm corr} &=& \tau_{\rm corr} n.
\label{Eq:Tcorrseq}
\end{eqnarray}
Here $\tau_{\rm pred}$ is the time to predict a single particle, $\tau_{\rm force}$ is the time to calculate the forces between two particles, and $\tau_{\rm corr}$ is the time spent to correct a single particle. The values for $\tau_{\rm pred}$, $\tau_{\rm force}$ and $\tau_{\rm
corr}$ have been measured by using a sample N-body simulation with
32768 particles, and are given in table\,\ref{Table:Netork} for the
various nodes in the G3. For a more practical comparison, we also
measured the compute speed (in floating point operations per second)
for each of the nodes. These measurements were carried out using the
Whetstone benchmark \citep{H.J.Curnow01011976}.

\begin{table}
\caption{ Machine performance specification and machine-specific
        constants. The first two columns show the name of the machine,
        followed by the country of residence. The third column
        indicates machine speed in Mflops, using the Whetstone
        benchmark. The last three columns give the time required for
        the CPU to perform one particle prediction ($\tau_{\rm
        pred}$), the time required for one force calculation between
        two particles ($\tau_{\rm force}$) and the time required for
        correcting one particle ($\tau_{\rm corr}$) respectively, all
        in microseconds.  }
\begin{tabular}{llcccc}
\hline
name&location&speed &$\tau_{\rm pred}$&$\tau_{\rm force}$&$\tau_{\rm corr}$\\
    &        &Mflops&[$\mu$s]         &[$\mu$s]          &[$\mu$s]\\
\hline
vader&NL     &$377 $&$0.247$          &$0.216$           &$4.81$ \\
palpatine&NL &$422 $&$0.273$          &$0.193$           &$2.39$ \\
yoda&JP      &$436 $&$0.131$          &$0.110$           &$1.29$ \\
skywalker&JP &$436 $&$0.131$          &$0.110$           &$1.29$ \\
obi-wan&US   &$1191$&$0.098$          &$0.148$           &$1.14$ \\
\hline
\label{Table:Netork}
\end{tabular}
\end{table}

\subsection{Grid of PCs with copy algorithm}

The performance model for a single PC can be extended to include the
parallel operation in the copy algorithm. In the copy algorithm, each process has a full copy of the system of $N$ particles, but only computes the active particles in a specific subset of $N/p$ particles. The result of this computation is sent to all other processes. We assume that all $p$ processes have comparable speed, and every process has an equally sized subset of $n/p$ active particles. For the copy algorithm, the host computation time ($T_{\rm
integrate}$) also consists of the time spent to communicate between processes ($T_{\tt MPI}$). Therefore,
\begin{eqnarray}
  T_{\rm integrate} &=& \sum^{\nsteps}_{i=1} 
                      \left( t_{\rm pred} + t_{\rm force} + t_{\rm corr}
		      \right) + T_{\tt MPI},
\label{Eq:Tintpar}
\end{eqnarray}
A process computes forces for its subset of $n/p$ active particles, and corrects only these particles. Therefore, a process
requires at most $N n/p$ force calculations per time step
\begin{eqnarray}
  t_{\rm force}&=&\tau_{\rm force} N \frac{n}{p},
\end{eqnarray}
and a process corrects at most $n/p$ particles, of which the time spent is given by
\begin{eqnarray}
  t_{\rm corr} &=& \tau_{\rm corr} \frac{n}{p}.
\label{Eq:Tcorr}
\end{eqnarray}
In a parallel system, time is spent not only on integrating the system ($T_{\rm integrate}$), but also on exchanging messages between processes ($T_{\tt MPI}$). This time is obtained by adding the time spent on overcoming network latency ($t_{\rm latency}$) and the time spent transferring particles ($t_{\rm bandwidth}$)
\begin{eqnarray}
  T_{\tt MPI} &=& \sum^{\nsteps}_{i=1} 
                     \left( 
                       t_{\rm latency} + t_{\rm bandwidth} 
		     \right) .
\label{Eq:Tmpi}
\end{eqnarray}
In our implementation $t_{\rm latency}$ is given by the sum of the latencies
of each MPI call in the code.  The copy algorithm uses 1
{\tt MPI\_Allgather} and 1 {\tt MPI\_Allgatherv} command (of which the
latencies both scale with $\log_2{p}$
\citep{DBLP:journals/pc/GualandrisZT07}) per block time-step,
resulting in a total time spent on latency of
\begin{eqnarray}
  t_{\rm latency} &=& (l_{\tt MPI\_Allgather} + 
                      l_{\tt MPI\_Allgatherv}) \log_2{p}.
\end{eqnarray}
The time used for transferring particle data is given by $t_{\rm
bandwidth}$, which is obtained by taking the total size of the data
that has to be communicated (which is assumed to scale with $2 (p-1)$
for all-to-all communications), and dividing it by the network
bandwidth ($\tau_{\rm bw}$). Particles are stored in 58 byte data
structures, resulting in
\begin{eqnarray}
  t_{\rm bandwidth} &=& \frac{n 116 (p-1)}{p\tau_{\rm bw}}.
\end{eqnarray}
For a wide area computer $t_{\rm latency}$ and $t_{\rm bandwidth}$ may
be quite substantial, but the separation in parts, as given
here, enables us to optimize our network computer with respect to the
communication characteristics.

\subsection{Grid of PCs with ring algorithm}

Unlike the copy algorithm, the ring algorithm (discussed in detail in
\citet{43389,83774}) does not use a single all-to-all communication
operation, and only requires the processes to have a partial copy of 
size $N/p$ of the system. Communication occurs in a a total of $p$ steps 
(or shifts). During every shift, each process performs a partial force integration by
calculating the forces between their local subset of $n/p$ active
particles and the $N/p$ particles stored in local memory. Then, each process sends its updated particles to their neighbor. 

To model the performance for the ring algorithm, we use the model for the copy algorithm and redefine the time spent on force calculations ($T_{\rm force}$) and the time spent to communicate between processes ($t_{\tt latency}$ and $t_{\tt bandwidth}$), as the force calculation and the MPI communications occur in multiple shifts. The time spent on a partial force calculation is given by
\begin{eqnarray}
  t_{\rm force, 1shift} &=& \tau_{\rm force} \frac{n N}{p^2}.
\end{eqnarray}
The total time spent on the force calculation is
given by the time for a partial force calculation ($t_{\rm force,
1shift}$) multiplied by the number of shifts ($p$),
\begin{eqnarray}
  t_{\rm force} &=& \tau_{\rm force} n \frac{N}{p}.
\end{eqnarray}
For every time step, our implementation of the ring algorithm uses 2 {\tt MPI\_Allreduce}
communication commands for initialization, and 1 {\tt MPI\_Sendrecv} operation for each shift. The time spent 
overcoming network latency is then
\begin{eqnarray}
  t_{\rm latency} &=& 2 \log_2{p} l_{\tt MPI\_Allreduce} + p l_{\tt MPI\_SendRecv}.
\end{eqnarray}
The ring algorithm is more bandwidth intensive than the copy algorithm, 
as all block subsets are sent and received during a ring shift, multiplying the
time spent on transferring particles by $2p$. Per particle, 58 bytes have 
to be transferred, therefore the time spent on transferring particles is given by
\begin{eqnarray}
  t_{\rm bandwidth} &=& 116 n /\tau_{\rm bw}.
\end{eqnarray}

\subsection{Single PC with GRAPE}\label{Sect:GRAPE}

The GRAPE-6Af is a dedicated hardware component developed by a group
of researchers led by Junichiro Makino at the University of Tokyo
\citep{2005PASJ...57.1009F}. The GRAPE-6Af is the smallest
commercially available GRAPE configuration, consisting of a single
GRAPE module with a peak speed of about 123 Gflops. It calculates the forces between particles, which is the bottleneck in the calculation, whereas the particle predictions and corrections are still mostly done on the host PC.

When a GRAPE is used, an $N$-body simulation over one $N$-body time unit consists of the following steps:
\begin{enumerate}
\item Read the input snapshot and initialize the $N$-body system.
\item Compute the next system time $t$ and select the block of active particles $n$ in the system.
\item Predict the positions and velocities of the $n$ active particles on the pc, and send the predicted values and the next system time to the GRAPE.
\item Predict the other particles in the system on the GRAPE.
\item Calculate the forces and their time derivatives, using the GRAPE, between the $n$ active particles and all $N$ particles in the system.
\item Retrieve the forces and their time derivatives from the GRAPE. 
\item Correct the positions, velocities and velocity derivatives of the $n$ active particles, and update their time steps.
\item Repeat from step 2 until $t$ has exceeded one $N$-body time unit.
\item Write the output of the simulation and terminate it.
\end{enumerate}
When using GRAPE
\begin{equation}
  t_{\rm pc} = t_{\rm pred} + t_{\rm corr},
\label{Eq:thost}
\end{equation}
where $t_{\rm pred} = n \tau_{\rm pred}$. 
The time spent to integrate particles for one $N$-body time unit is given by
\begin{eqnarray}
  T_{\rm integrate}&=&\sum^{\nsteps}_{i=1} 
                   \left( 
                     t_{\rm pc} + t_{\rm grape} + t_{\rm comm}
		   \right).
\end{eqnarray}
Here 
\begin{eqnarray}
  t_{\rm grape}&=& \tau_{\rm pipe} n N,
\end{eqnarray}
is the time for calculating the forces on the GRAPE. The time needed by the GRAPE to calculate the force between two particles is given by $\tau_{\rm pipe}$.  The communication between host and GRAPE is given by \citep{2005PASJ...57.1009F}
\begin{eqnarray}
  t_{\rm comm}&=&60 t_i n + 56 t_f n + 72 t_j n.
\end{eqnarray}
Here the time to respectively send or receive 1 byte of data to the
GRAPE during different steps is given by $t_i$, $t_n$ and $t_j$,
respectively. During these steps respectively 60, 56 and 72 bytes per
particle in the block are transferred. We assume that $t_i = t_f =
t_j$. We derive $t_j$ by measuring $\tau_{\rm Gsend}$, which is the time to
send one 72-byte particle to the GRAPE. Therefore, $t_j = \tau_{\rm
Gsend}/72$. By rewriting $t_i$, $t_n$ and $t_j$ as factors of
$\tau_{\rm Gsend}$, we can simplify the equation for $t_{\rm comm}$ to
\begin{eqnarray}
  t_{\rm comm}&=&\left(60+56+72\right) \left( \tau_{\rm Gsend}/72 \right) n.
\end{eqnarray}
Time spent on calculating forces on the GRAPE ($t_{\rm grape}$) cannot be
directly measured by timing parts of the code, because the GRAPE
force calculation includes some communication between host and GRAPE
as well. However, we can derive $\tau_{\rm pipe}$ from the
total time of the force calculation as was done in
\cite{2007NewA...12..357H}. Therefore, we can rewrite $\tau_{\rm pipe}$ as
\begin{eqnarray}
  \tau_{\rm pipe} &=& \frac{1}{N} 
		      \lbrack
                         t_{\rm force} - \left( 116 \tau_{\rm Gsend}/72 
		                         \right) 
		      \rbrack .
\end{eqnarray}
The time spent on performing $N$ force calculations $t_{\rm force}$ 
is given by,
\begin{eqnarray}
t_{\rm force}&=& \tau_{\rm Gforce} N,
\end{eqnarray}
where $\tau_{\rm Gforce}$ is the time spent to calculate forces between two particles. We then introduce the time constant ($\tau_{\rm Gforce}$) in the function for $\tau_{\rm pipe}$,
\begin{eqnarray}
  \tau_{\rm pipe}&=& \tau_{\rm Gforce} - \left( \left( 116 \tau_{\rm Gsend}/72\right) \frac{1}{N} \right).
\end{eqnarray}
Using our derived functions for $t_{\rm grape}$ and $t_{\rm comm}$, we
are now able to model the performance of the GRAPE. As mentioned in
\citep{2005PASJ...57.1009F}, $\tau_{\rm Gforce} \approx 4.3 \cdot 10^{-10}$ s.

\subsection{Grid of PCs with GRAPE and copy algorithm}


When using GRAPE and a parallel algorithm, the time spent to integrate particles for one $N$-body time unit is given by
\begin{eqnarray}
  T_{\rm integrate}&=&\sum^{\nsteps}_{i=1} 
                   \left( 
                     t_{\rm pc} + t_{\rm grape} + t_{\rm comm}
		   \right) + T_{\tt MPI}.
\end{eqnarray}

We determine time spent communicating between hosts ($T_{\tt MPI}$) using Eq.\,\ref{Eq:Tmpi}. To determine the time spent on the host ($t_{\rm pc}$) we use the equation for the single process with GRAPE (see Eq.\,\ref{Eq:thost}). However, as we correct only $n/p$ particles in parallel algorithms, we apply Eq.\,\ref{Eq:Tcorr} to determine the time spent by the process on correcting particles.

In a parallel setup of GRAPEs, each process needs to
communicate and calculate forces for a subset of $n/p$ particle in
every block. We replace $n$ by $n/p$ in our equations
for $t_{\rm grape}$ as well as $t_{\rm comm}$. Therefore, the time spent on
calculating forces is given by
\begin{eqnarray}
  t_{\rm grape} &=& \lbrack N \tau_{\rm Gforce} - \left(116 \tau_{\rm Gsend}/72 \right)
                    \rbrack \frac{n}{p},
\label{Eq:tgrape}
\end{eqnarray}
and communication between between the hosts and the GRAPEs becomes, 
\begin{eqnarray}
  t_{\rm comm}&=& \left( 60+56+72 \right) \left( \tau_{\rm Gsend}/72
                   \right) \frac{n}{p}.
\end{eqnarray}

\subsection{Grid of PCs with GRAPE and ring algorithm}

In a ring algorithm, each process computes the forces between a local set of $n/p$ active particles and the local system of $N/p$ particles during a shift.
Then, it sends the results to its next neighbor and receives another $n/p$ particles from its other neighbor. Each of the nodes spends $t_{\rm grape,
1shift}$ calculating the forces for $n/p$ particles during one shift. Before the node has integrated the force on all $n$
particles, a total of $p$ shifts have passed, resulting in a total
compute time for this node of
\begin{equation}
  t_{\rm grape} = p t_{\rm grape, 1shift},
\end{equation}
where the time to calculate forces for a single shift ($t_{\rm grape, 1shift}$) is
\begin{eqnarray}
  t_{\rm grape, 1shift} &=& 
                    \lbrack 
                            \left(N \tau_{\rm Gforce}/p\right) - 
                            \left( 116 \tau_{\rm Gsend}/72 \right)
                    \rbrack 
                    \frac{n}{p}.
\end{eqnarray}
When GRAPE is used, particles are 74 bytes each because they contain
two additional arrays for storing the old acceleration and old jerk.
Due to this increased particle size,
\begin{eqnarray}
  t_{\rm bandwidth} &=& 148 n/\tau_{\rm bw},
\end{eqnarray}
whereas $t_{\rm latency}$ remains unchanged.  The time spent
communicating between hosts ($T_{\tt MPI}$) is calculated as for the ring algorithm without GRAPE.

\section{Results of the performance model}\label{Sect:ModelResults}

We have applied the performance model from the previous section to the
results presented in \S\,\ref{Sect:Results}.  In Fig.\,\ref{Fig:Copy}
we compare the measured wall-clock time ($T_{\rm app}$) for the copy
algorithm on the grid with the performance model, Fig.\,\ref{Fig:Ring}
shows a similar comparison for the ring algorithm.  To guide the eye,
the results for a single GRAPE are also presented in both figures. The
performance model tracks the real measurements quite satisfactorily,
giving a slightly lower computation time for a single GRAPE
while giving a slightly higher computation time for a simulation across
grid sites.

The communication overhead of a distributed computer often
renders high performance computing on a grid inefficient. However, in
the N-body problem the compute time scales with $N^2$ whereas the
communication scales linearly with $N$. For sufficiently large $N$,
there will eventually be a point where relatively little time is lost
communicating, and the compute resources are efficiently used.

In figures\,\ref{Fig:Copy} and \ref{Fig:Ring} we can see that, for
GRAPE-enabled simulations, break-even between calculation and
communication is reached around $N \simeq 10^6$. For large $N$, a grid
of two GRAPEs will outperform a single GRAPE. Our grid setup included
three GRAPE-enabled sites. The location of these sites (Asia, Europe
and America) were as widely distributed as physically possible. A more
modest grid across a single continent, will perform considerably
better than a global grid.  With the performance model that we
constructed in \S\,\ref{Sect:model}, we can now study various grid
topologies without the need to physically build the environment and
create a virtual organisation.

\subsection{Future Prospects}

We applied the performance model to three hypothetical grids of GRAPE
nodes.  These three grids are: 1) a grid of all the available
GRAPEs on the planet, 2) a grid of sites with more than 1 Tflops
in GRAPE speed, and 3) a recently established Dutch grid (Dutch
ASCII Computer, DAS-3\footnote{http://www.starplane.org/das3/})
equipped with GRAPEs.

Since the launch of GRAPE-6, a total of 1115 GRAPE-6 modules have been
deployed worldwide.  Japan leads the GRAPE-yard with more than 800
modules, followed by the US (119) and Germany (62).  At the moment
there are 876 GRAPE-6 modules in Asia, 132 in North America and 107 in
Europe.  In Table \,\ref{Tab:G3_university} we list the sites with
more than 1 Tflops peak-performance in GRAPE hardware and their
network characteristics. The network roundtrip with the longest
latency is a roundtrip between Japan and Ukraine, whereas the network
link with the longest latency (not given in Table
\,\ref{Tab:G3_university}) is the transpacific line between Japan and
the US.

\begin{table}
\caption{Overview of the major GRAPE clusters ($> 1$Tflops) on
         planet Earth, including their relative network latency
         characteristics.  The first column identifies the site,
         followed by the name of the institute, the country and the
         number of GRAPE modules (8 modules provide $\sim 1$ Tflops
         peak performance.).  The fifth column identifies the site
         with which the latency, given in column \#6, is shortest.
         The one but last column (column \#7) identifies the site with
         which the latency, given in column \#8, is longest.  The
         total number of GRAPE modules is 996.
\label{Tab:G3_university}
}

\begin{tabular}{llll|clcl}
\hline
ID&institution            &country    &\# GRAPEs &\multicolumn{2}{c}{nearest site} &\multicolumn{2}{c}{farthest site}\\
 &                        &           &     &  &[ms]&  & [ms] \\
\hline
A&Nat. Astronomical Obs.  &Japan      &576  &B & 2  &M & 330       \\
B&Tsukuba University      &Japan      &240  &A & 2  &M & 330       \\
C&AMNH                    &US         &40   &G & 5  &B & 190       \\
D&Astron. Rechen Inst.    &Germany    &30   &K & 2  &B & 280       \\
E&Rochester Institute     &US         &26   &F & 5  &B & 170       \\
F&McMaster University     &Canada     &13   &E & 5  &B & 170       \\
G&Drexel University       &US         &12   &C & 5  &B & 190       \\
H&Max Planck Institute    &Germany    &12   &D & 10 &B & 290       \\
I&University of Amsterdam &Netherlands&10   &K & 7  &B & 266       \\
J&Wien University         &Austria    &10   &H & 10 &B & 290       \\
K&Bonn University         &Germany    &9    &D & 2  &B & 280       \\
L&Cambridge University    &England    &9    &I & 10 &B & 260       \\
M&Main Astronomical Obs.  &Ukraine    &9    &J & 30 &B & 330       \\
\hline 
\end{tabular}
\end{table}

Organizing all the GRAPEs on the planet would be a challenging
political problem. Organizing only the 13 largest sites would be
somewhat easier, therefore we included a performance prediction of
such an infrastructure as well. Constructing a virtual organization
within the 4 universities (University of Amsterdam, Free University of
Amsterdam, Leiden University and Delft Technical University) that
participate in the DAS-3 project (and equipping the 270 available
DAS-3 nodes with specialized hardware) would be much easier than doing
this across various countries. The Dutch DAS-3 Grid is equipped with a
fast Myrinet-10G internet and distributed across the Netherlands,
connected by 10Gb light paths between clusters. The latency for the
longest path is estimated to be 3\,ms, and we estimate the bandwidth 
of the connections to be 0.5\,GB/s.

In the sequential case we do not take memory limitations into
account. This assumption is unrealistic for predicting the performance
of a GRAPE, since $N<262144$, but GPUs have a similar
performance to GRAPE, and are able to store up to 13 million particles
\citep{2007astro.ph..2058P}. In late 2007, the launch of a
double-precision GPU is expected, making GPUs usable for
production-type direct-method N-body simulations. Additionally, in
recent years the amount of memory on GPUs has been steadily
increasing, and we expect this trend to persist in the near future.

We model the ring algorithm on both the G3 and on the 13 largest sites
(see Table \,\ref{Tab:G3_university}).  We adopted {\em palpatine}
(see Table \,\ref{Tab:G3}) as the workhorse host for the GRAPEs and
adopted the network characteristics as listed in Table
\,\ref{Tab:G3_university}. The ring algorithm for our hypothetical
grid experiment was assumed to be optimized for the use of
distributed clusters of GRAPEs. The algorithm avoids latency intensive
network links by combining communication of local clusters from
multiple shifts. Thus, communication for the local (across node)
network is separated from the global (across sites)
communication. Finally, we assume that all networks between the sites
have reasonable support for MPI multicast and gather operations, and
that the latency of these operations scales with $\log_2{p}$

\begin{figure}
  \centering
  \includegraphics[angle=270,scale=0.4]{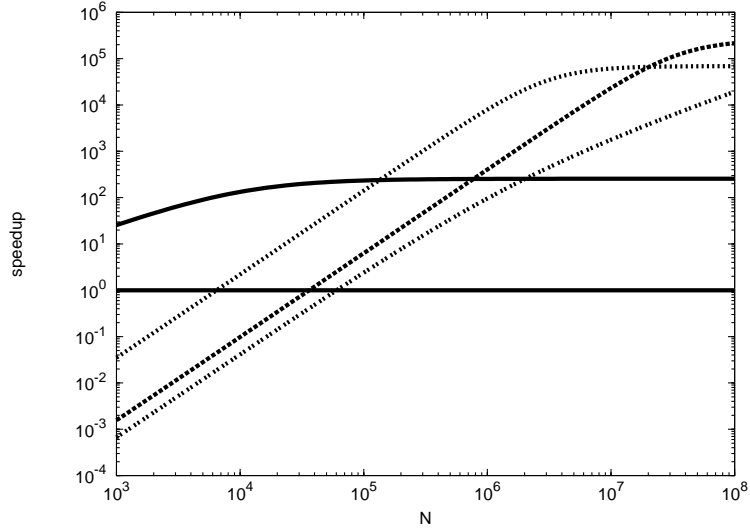}
  \caption{ Speedup prediction of possible GRAPE grid setups compared
to a single CPU. The solid lines indicates execution time using 1 CPU
(horizontal reference line) or 1 GRAPE with infinite memory (curved
line). The double-dotted line indicates the predicted speedup if all
1115 GRAPEs are linked together to perform one simulation using an
optimized ring algorithm. The dashed line indicates the predicted
speedup if all GRAPE sites with more than 1 Tflops are linked together
to perform one simulation using an optimized ring algorithm. The
dotted line indicates the speedup if all 270 nodes in the Dutch DAS-3
grid would be equipped with GRAPEs.
  \label{Fig:PredSpeedup}
  }

\end{figure}

The results of the hypothetical global GRAPE grid are presented in
Fig.\,\ref{Fig:PredSpeedup}.  Here we see that a global grid in which
all GRAPEs participate outperforms a single GRAPE by about two orders of
magnitude for $N \apgt 10^8$ particles.  For a sufficiently large
number of particles ($N\apgt 10^9$) the total peak performance of the
global GRAPE grid approaches about 75\,Tflops. Eventually, the grid
with all the GRAPES would outperform the grid with only the largest
machines by about 25\%, proportional to the number of GRAPEs in the
two setups.  When running simulations of $N\sim 10^6$ it is faster to
run on three large GRAPE sites, than to use all the GRAPEs on the
planet in parallel.

The dashed curve in Fig.\,\ref{Fig:PredSpeedup} shows the performance
of the model assuming that all the 270 nodes of the DAS-3 were
equipped with GRAPE-6Af hardware. With such a setup, the maximum
performance of about 35\,Tflops is achieved for $N\sim 10^7$
particles. This is an interesting number for production simulations
for astronomical research.

\begin{figure}
  \centering
  \includegraphics[angle=270,scale=0.4]{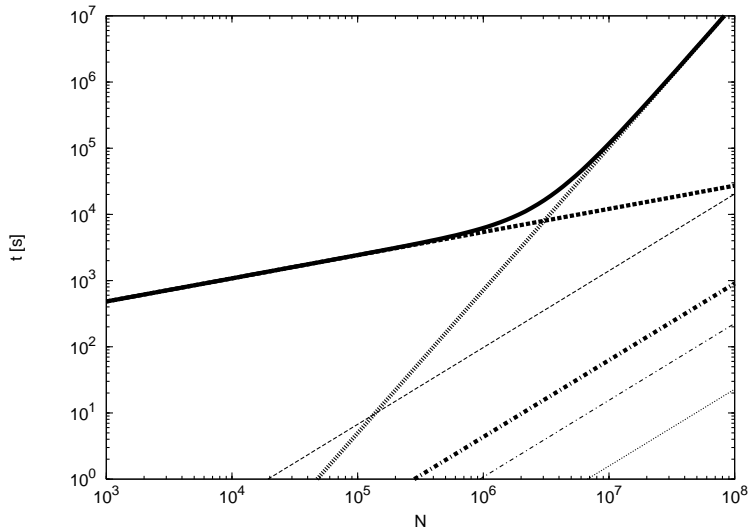}    
  \caption{ Predicted decomposition of performance of a DAS-3 GRAPE
grid. The thick solid line indicates total execution time and the flat
thick dashed line indicates time spent due to network latency. The
thin dashed line indicates time spent on using the network bandwidth
and the steep thick dotted line indicates time spent on calculating
forces. The three bottom lines indicate time spent on communication
between hosts and GRAPEs (upper thick dash-dotted line), correcting
particles (middle dash-dotted line), and predicting particles (lower
dotted line)
  \label{Fig:DAS3}
  }
\end{figure}

In Fig.\,\ref{Fig:DAS3} we present the wall-clock time for each of the
different ingredients of a grid calculation with GRAPEs on the DAS-3,
using the performance model.  Break-even between calculation (straight
solid curve) and communication (thick dotted curve) is achieved around
$N\sim 3 \cdot 10^6$. For this large number of particles the communication
between GRAPE and host, the predictor and the corrector steps require
little CPU time compared to the force calculation on the GRAPE.  For
$N \apgt 6 \cdot 10^6$ this setup would give an efficient use of the special
processors, and high performance calculations on the grid would then
be quite efficient.

\section{Discussion and Conclusions}

We studied the potential use of a virtual organization in which GRAPEs
are used in a wide area grid. For this purpose, we developed a
performance model to simulate the behavior of a grid in which each of
the nodes is equipped with special purpose GRAPE hardware.  We tested
the performance model with an actual grid across three sites, each of
which is located on a different continent. We used GRAPE hardware in
Japan, the Netherlands and the USA simultaneously for calculations of
1024 up to 196608 particles. 

With these particle numbers we were able to have a better performance
than a single computer without GRAPE. We measured a grid speedup
of $\Gamma \sim 1.37$ for a grid of PCs, and a
grid of GRAPEs performs another $\sim 4$ times faster. On the entire
range of $N$ we were unable to reach superior speed compared to a
single GRAPE. However, we estimate that a small intercontinental grid
of GRAPEs will reach superior performance for $N \apgt 3 \cdot 10^6$
particles.

We used our grid calculations with GRAPE to construct and calibrate a
performance model, with which we studied the performance of a
world-wide grid of GRAPEs.  When all the GRAPEs on the planet would
participate in a virtual organization it is possible to utilize the
total machine's performance, but only for really large systems of
$N\apgt 10^9$. Though the total performance for such a setup would be
about 75\,Tflops, such large $N$ would still be impractical to run
for production astronomical simulations.

We conclude that organizing all the major GRAPEs on the planet in a
virtual organization is probably not worth the effort.  Organizing a
few of the largest sites with GRAPEs within one continent, however,
appears politically doable and computationally favorable.  For the
DAS-3, for example, the GRAPEs would be used at maximum performance
for a feasible number of stars. Modern simulations of up to about a
million stars have been done before using GRAPE
\citep{2004Natur.428..724P}, but these calculations were performed on
a single cluster, rather than on a grid. A grid setup as proposed
here would allow the simulation of a few million stars within a
reasonable time span.

If we were to equip the full DAS-3 wide area computer in the Netherlands
with GRAPEs, maximum performance would already be achieved for $N\sim
6 \cdot 10^6$ particles. Though still large, such simulations would be
very doable and have practical applications. We estimate that running
a system of $N = 10^6$ stars with a Salpeter mass spectrum
\citep{1955ApJ...121..161S} over a wide range of stellar masses to the
moment of core collapse would take about 4 months. The simulation would
still be mostly dominated by network latency, but the high-throughput 
networking in the DAS-3 completely removes the bandwidth bottleneck.

We have mainly discussed the use of GRAPEs in a virtual organization,
but new developments in using graphical processing units appear to
achieve similar speeds as GRAPEs
\citep{2007astro.ph..2058P,2007astro.ph..3100H,2007arXiv0707.0438B}. In
addition, GPUs are equipped with a larger amount of memory, which
allows us to exploit more memory-intensive, but also faster, parallel
algorithms. Future grids are likely to be equipped with GPUs, as the
GPU will become part of the standard equipment for every PC.

Although our proof-of-concept infrastructure was of limited size, we
have shown that it is possible to use dedicated hardware components
located across clusters for high-performance computing. Though the
current performance over globally connected grids leaves a lot to be
desired and much optimization remains to be done, the concept of using
dedicated hardware components worldwide in parallel has been shown to
work. It can therefore be applied to solving individual
tightly-coupled scientific problems or as ingredient of a complex
multi-physics simulation, such as simulating a full galaxy, given that the
problem size is sufficiently large to overcome the networking
limitations.

\section{Acknowledgements}
We are grateful to Mary Inaba and Cees de Laat for discussion and
support in realizing this grid setup, and to Alessia Gualandris for
providing some of the codes and feedback on this work. We are also
grateful to Alfons Hoekstra, Marian Bubak and Stefan Harfst for
fruitful discussions on the contents of this paper. This research is
supported by the Netherlands organization for Scientific research
(NWO) grant \#643.200.503 and by the European Commission grant for the
QosCosGrid project (grant number: FP6-2005-IST-5 033883), and we thank SARA
computing and networking services, Amsterdam for technical support.

\bibliographystyle{elsart-harv}

\bibliography{GPZGMcM2008}

\end{document}